\documentclass[conference]{IEEEtran}
\IEEEoverridecommandlockouts
\usepackage{cite}
\usepackage{amsmath,amssymb,amsfonts}
\usepackage{algorithmic}
\usepackage{graphicx}
\usepackage{textcomp}
\usepackage{xcolor}
\usepackage{mathrsfs}
\usepackage{caption}
\usepackage{subcaption}
\usepackage{wrapfig}
\usepackage{enumitem}

\usepackage{eqparbox}
\usepackage{float}
\usepackage{lipsum}
\usepackage{multirow}
\usepackage{scalerel}
\usepackage{hyperref}
\usepackage{mathtools}
\usepackage[capitalize]{cleveref}
\Crefname{chapter}{Chap.}{Chaps.}
\Crefname{section}{Sec.}{Secs.}
\Crefname{figure}{Fig.}{Figs.}
\usepackage{titlesec}
\titlespacing*{\section}{0pt}{0.1\baselineskip}{0.1\baselineskip}
\titlespacing*{\subsection}{0pt}{0.1\baselineskip}{0.2\baselineskip}
\titlespacing*{\subsubsection}{0pt}{0.1\baselineskip}{0.2\baselineskip}

\makeatletter

\g@addto@macro\normalsize{%
	\setlength\abovedisplayskip{1pt plus 0pt minus 1pt}
	\setlength\belowdisplayskip{1pt plus01pt minus 1pt}
	\setlength\abovedisplayshortskip{2pt plus 1pt minus 1pt}
	\setlength\belowdisplayshortskip{2pt plus 1pt minus 1pt}
}
\makeatother

\def\amm{\theta_{\scaleto{\text{AM }}{4pt}}}

\def\leftctx{a_{{s_t}-1}}
\def\center{a_{s_t}}
\def\rightctx{a_{{s_t}+1}}

\newcommand{\argmax}{\operatornamewithlimits{argmax}}

\DeclareMathSizes{6}{6}{7}{7}

\def\BibTeX{{\rm B\kern-.05em{\sc i\kern-.025em b}\kern-.08em
    T\kern-.1667em\lower.7ex\hbox{E}\kern-.125emX}}

\begin{document}

\title{Right Label Context in End-to-End Training of Time-Synchronous ASR Models}

\author{\IEEEauthorblockN{Tina Raissi{$^1$}, Ralf Schl\"uter{$^{1,2}$}, Hermann Ney{$^{1,2}$}}
\IEEEauthorblockA{\textit{ $^1$Machine Learning and Human Language Technology Group, RWTH Aachen University} \\
\textit{\textit{
		$^2$AppTek GmbH}}\\
Germany \\
}

}

\maketitle

\begin{abstract}
	Current time-synchronous sequence-to-sequence automatic speech recognition~(ASR) models are trained by using sequence level cross-entropy that  sums over all alignments.\ Due to the discriminative formulation, incorporating the right label context into the training criterion’s gradient causes normalization problems and is not mathematically well-defined.\ The classic hybrid neural network hidden Markov model~(NN-HMM) with its inherent generative formulation enables conditioning on the right label context.\ However, due to the HMM state-tying the identity of the right label context is never modeled explicitly.\ In this work, we propose a factored loss with auxiliary left and right label contexts that sums over all alignments.\ We show that the inclusion of the right label context is particularly beneficial when training data resources are limited.\ Moreover, we also show that it is possible to build a factored hybrid HMM system by relying exclusively on the full-sum criterion. Experiments were conducted on Switchboard 300h and LibriSpeech 960h.
\end{abstract}

\section{Introduction}
The training of time-synchronous automatic speech recognition~(ASR) models requires a set of possible alignment paths between the input feature and output label sequences, following a certain label topology~\cite{graves2006connectionist,graves2012sequence,tripathi2019monotonic,Raissi+Zhou+:2023}.\ The common sequence level cross-entropy training criterion that uses the sum over all alignments can be replaced by a frame-wise cross-entropy on the single best alignment path without significant loss of performance.\ This approach is the prevalent technique for training Gaussian mixture hidden Markov model (GM-HMM) and later the hybrid neural network~(NN)-HMMs~\cite{Bourlard+Morgan:1993}.\ In these frameworks, after the initial linear segmentation, the alignment path is iteratively refined through multiple realignment steps.\ This type of training is designed to address convergence issues that occur when using the sum over all alignment paths~\cite{Raissi+Zhou+:2023}.\ Recent advances in encoder architecture design leveraging global dependencies through the self-attention mechanism and recurrent layers on the feature input side, along with the use of downsampling and larger label output units such as wordpieces and byte-pair encoding units~\cite{sennrich-etal-2016-neural} have mitigated part of this convergence issue.\ The mentioned modeling choices reduce the number of alignment paths, thereby imposing fewer constraints on the alignment quality.\ 

Another difference between the classic and the sequence-to-sequence~(seq2seq) frameworks is the use of the output label context in the frame-wise probability distribution.\ The autoregressive factorization in seq2seq models that use label context does not allow for the inclusion of right (future) context. Whereas, in principle, there is no such constraint on right label context modeling in the classic generative approach.\ Nevertheless, even the right label context modeled via generalized triphone states in classic approaches necessitates a parameter tying with several inconsistencies in the resulting statistical model~\cite{bell2016multitask,raissi2020fh}.\ Moreover, it has been shown that models such as recurrent neural network transducers~(RNN-T)~\cite{graves2012sequence} do not require the full label history to achieve full performance.\ The label context history in these models can be restricted to one or two previous output labels with no word error rate~(WER) degradation~\cite{ghodsi2020rnn,zipformer,zhou:phoneme-transducer:2021,gnat}.\ 

 The issues investigated in this work are two-fold: first, whether the inclusion of the right label context in the model definition leads to ASR performance gain with full-sum training, and second, whether it is possible to build an end-to-end hybrid HMM system.\ For the first question we expanded on an initial investigation~\cite{raissi21:tchham} with three key extensions and differences: (1) we evaluate an ASR model trained from scratch and with no intermediate forced alignment step, (2) we omit the HMM state prior during training, and (3) we use the Conformer encoder architecture~\cite{Gulati+Qin+:2020conformer}.\ We show that the inclusion of the right label context is particularly beneficial when data resources are limited.\ Moreover, we build a factored hybrid HMM~\cite{raissi2020fh,raissi2023competitive} using only the full-sum criterion, and achieve comparable performance to a multi-stage pipeline system that relies on an external alignment.\ To the best of our knowledge, no prior approach in the literature has explicitly modeled the right label context for full-sum training.\

\section{Training Criteria and Inference Rules}
\label{sec:criteriarule}

\subsection{Modeling Approach}
\label{subsec:modeling}
The standard Bayes decision rule for automatic speech recognition for an input feature sequence $X$ and an output word sequence $W$ is defined in \cref{eq:bayes1}.\
\begin{align}
	\label{eq:bayes1} \footnotesize
	X \rightarrow\tilde{W}(X) = \underset{W}{\argmax} \left\lbrace P_{_{\theta}}(W | X)  \right\rbrace \phantom{0000000000} \\
	\label{eq:bayes2} 
	\phantom{X \rightarrow\tilde{W}(X)}= \underset{W}{\argmax} \left\lbrace P_{_{\theta_{\scaleto{\text{AM}}{2pt}}}} (X | W) \cdot P_{_{\theta_{\scaleto{\text{LM}}{2pt}}}}(W) \right\rbrace 
\end{align}
In classic generative approaches the acoustic model (AM) and the language model (LM) are treated as distinct components, as indicated in \cref{eq:bayes2}. In contrast, modern seq2seq models, commonly referred to as end-to-end approaches, eliminate this separation by integrating the acoustic and language modeling into a single, unified framework, as in \cref{eq:bayes1}.\ 

The training criterion that determines the optimal parameters $\theta$ and $\amm$ for the acoustic models in the two approaches makes use of a frame-level probability distribution that either generates an input feature at time frame $t$ or predicts a label conditioned on the input and, optionally, a label history.\ The input and output at each time frame are matched via an alignment or path through the lattice of all possible alignments between the two feature and label sequences.\ Given a phoneme sequence $\phi$ corresponding to $W$, the marginalization over all alignments is done by using the hidden Markov sequence $S$ and the blank augmented sequence $Y$ both of length $T$.\ Denote $h_1^{T} = Enc(X)$ as the neural representation of the input acoustic feature sequence $X$ with downsampling by a factor of four for all our models.\ 
We focus only on \textit{time synchronous} modeling approaches where independent of the intermediate factorization or use of auxiliary factors, the final learned probability distribution used in Bayes rule for decoding is only conditioned on the feature input. Namely, a first order label context model proposed as (diphone) factored hybrid~\cite{bourlard1992cdnn,raissi2020fh}, and three zero order label context models: a connectionist temporal classification~\cite{graves2006connectionist}, a posterior HMM~\cite{raissi2022hmm}, and a hybrid NN-HMM trained with left and right auxiliary factors~\cite{raissi21:tchham}.\ The label set across all topologies consists in single state phonemes from the pronunciation lexicon, with the end-of-word (EOW) distinction.\ We consider a single pronunciation for each word.\ For the CTC model, an additional blank token is used, while only the HMM model explicitly utilizes a silence token.

\subsection{Right Label Context in Full-Sum training}
\label{subsec:rightcontext}
To discuss the integration of the right label context, we first perform the frame-level factorization under model specific assumptions for the full-sum training criterion.\ The frame-level factorization of CTC model can be seen in \cref{eq:decomposediscr}.\ 

\begin{footnotesize}\vspace{-0.2cm}
\begin{align}
	\label{eq:decomposediscr} \footnotesize
	\sum_{Y} P(\phi, Y | h_1^{T}) =  \sum_{y_1^{T}:\phi} \prod_{t=1}^T P(y_t | h_t)	
\end{align}
\end{footnotesize}
The frame-wise posterior of CTC  does not allow for the integration of any type of label context due to the independence assumption. \ Discriminative approaches such as RNN-T that do not drop the dependency on the label context rely on a factorization that yields merely a left label history, with arbitrary length, while excluding any right context.\ Access to the right context differs in the generative approach. For models following the HMM assumption, we first carry out the decomposition into the emission and transition probabilities as in \cref{eq:decomposegen1}. In neural based HMMs, \cref{eq:decomposegen2}  is derived by applying Bayes identity. 

\begin{footnotesize}\vspace{-0.3cm}
	\begin{subequations}
		\begin{align}
			\label{eq:decomposegen1} 
			\sum_{S} P(h_1^{T}, S | \phi)  &= \underset{s_1^T:\phi}{\sum} \prod_{t=1}^T P(h_t | s_t, \phi)P(s_t | s_{t-1}) \\
			\label{eq:decomposegen2} 
			& \sim  \underset{s_1^T:\phi}{\sum} \prod_{t=1}^T \frac{P(s_t, \phi | h_t)}{P_{\text{prior}}(s_t, \phi)}P(s_t | s_{t-1})
		\end{align}
	\end{subequations}
\end{footnotesize}
The additional term $P(h_t)$ is constant with respect to both the training criterion and the Bayes decision rule, and is therefore omitted here.\

In contrast to CTC, the emission probability in \cref{eq:decomposegen1} generates the input conditioned on the tuple $\{s_t, \phi\}$.\ The main emphasis of the analysis is the extent to which the HMM state $s$ at time frame $t$ has access to the sequence $\phi$.\ During training, the phoneme sequence $\phi$ is fully available, since we have access to the transcript and the pronunciation lexicon.\ For decoding, when using a finite state acceptor, the structure of the graph can allow the distinction between alignment states of a certain center phoneme in different contexts following the context composition C in the HCLG composition~\cite{mohri2002weighted}.\ We consider the augmentation of the phoneme aligned to the state $s_t$ at time frame $t$ with its right and left phonemes from the original phoneme sequence $\phi$.\ Let $a_{s_t}$ be a position aware labeling function that given the hidden Markov alignment state $s$ at time frame $t$ provides the label of the aligned phoneme within $\phi$.\ For simplicity the access to the adjacent phoneme labels is done via $\leftctx$ and $\rightctx$, for left and right phonemes, respectively.\ Explicit modeling of the left, center, and right phonemes follows the substitution of $\{s_t, \phi\}$ with $\{ \leftctx , a_{s_t},\rightctx \}$, yielding \cref{eq:triphone}.
\begin{equation}\vspace{-0.2cm}
	\label{eq:triphone} \footnotesize
 \sum_{S} P(h_1^{T}, S | \phi)  =   \underset{s_1^T:\phi}{\sum} \prod_{t=1}^T \frac{P(\leftctx , a_{s_t},\rightctx | h_t)}{P_{\text{prior}}(\leftctx, a_{s_t},\rightctx)}P(s_t | s_{t-1})
\end{equation}

\subsection{Decoding}
The best word sequence for all approaches is obtained via Viterbi decoding and by using a scaled log linear formulation, using prior, transition, and language model~(LM) scales $\beta$, $\eta$, and $\lambda$, respectively.\ The decoding of all hidden Markov models mentioned in \cref{subsec:modeling} relies on \cref{eq:phmmdecode}, except for the diphone model.\ This model uses $P(\leftctx, \center | h_t)$ and $P(\leftctx, \center)$ for the label posterior and priors, respectively~\cite{raissi2020fh}.\
\begin{equation} 
	\label{eq:phmmdecode} \footnotesize
	\underset{W}{\argmax} \hspace{1mm} \Big[ P^{\lambda}_{\text{LM}}(W) \underset{s_1^T:\phi}{\max} \prod_{t=1}^T \frac{P(a_{s_t} | h_t)}{P^{\beta}_{\text{prior}}(a_{s_t})} P^{\eta}(s_t, s_{t-1}) \Big] 
\end{equation}
A similar formulation is used also for CTC by omitting the transition probability~\cite{Raissi+Zhou+:2023}.
\subsection{Right Label Context Contribution to the Gradient}
Following the model definition of \cref{eq:triphone}, the gradient for the optimization of the log likelihood criterion with respect to the acoustic model parameters $\amm$ is

\begin{footnotesize}\vspace{-0.2cm}
\begin{subequations}
	\begin{align}\label{eq:gradient} 
		&\frac{\partial}{\partial \amm} LL(\amm) = \\
		 &\sum_{t,s} \left[  \gamma_t(a_{s-1}, a_{s}, a_{s+1}| h_1^T, \amm)  \frac{\partial}{\partial \amm} \log  \frac{P(a_{s-1} , a_{s},a_{s+1} | h_t)}{P_{\text{prior}}(a_{s-1} , a_{s},a_{s+1})} \right] \nonumber \\
		  &\approx \sum_{t} \sum_{a \in \{a_{s-1}, a_{s}, a_{s+1}\}} \left[  \gamma_t(a| h_1^T, \amm)  \frac{\partial}{\partial \amm} \log  \frac{P^{\alpha_a}(a | h_t)}{P^{\beta_a}_{\text{prior}}(a)} \right]	\label{eq:gradientsimplify}  
	\end{align}
\end{subequations}
\end{footnotesize}
Where $\gamma$ defined in \cref{gamma} guarantees the local normalization over the set of acoustic model labels at each time step and is calculated efficiently via forward-backward algorithm.\ 
\begin{equation} \label{gamma} \footnotesize
	\gamma_t(\leftctx, \center, \rightctx| h_1^T, \amm) = \frac{\underset{S:\phi ,s_t=s}{\sum}P(h_1^T, S|\phi, \amm)}{\underset{S:\phi}{\sum}P(h_1^T,S|\phi, \amm)}
\end{equation}
For efficient implementation of \cref{eq:gradient}, we apply two additional simplifying assumptions motivated later in \cref{subsec:simplify}: (1) elimination of the dependency to the label context for each factor, and (2) a state prior scale $\beta$ equal to zero.\ The resulting gradient after the application of the mentioned assumptions can be seen in \cref{eq:gradientsimplify}. 
\subsection{Role of Scales in Training}
\label{subsec:scales}
The standard solution for training high performance HMM based systems is to use an external alignment, while the novel seq2seq approaches such as CTC use the full-sum criterion.\ The optimization based on full-sum criterion is performed as a form of expectation-maximization algorithm.\ The training can begin with randomly initialized parameters, eliminating the need for an alignment between the speech input and the output label sequences.\ There are several aspects in the choice of common CTC approaches that play an important role for the convergence of the model.\ This includes the choice of output label unit, the acoustic input downsampling factor, and the label topology~\cite{zhao2023regarding, Raissi+Zhou+:2023}.\ For the standard CTC label topology, the blank label assumes both the role of continuation of the duration of the last emitted label, and the role of silence.\ Due to this ambiguity, the use of blank in the CTC label topology allows the model to have more freedom in the choice of the output label start and end boundaries.\ Given a segment, the label can be emitted at any arbitrary position.\ This is not the case for the HMM topology, where we have a partition that starts with the emission of the label and ends after the last label loop.\ This distinction results also in different convergence behaviors in the two models.\ When training a model from scratch using full-sum training, it is desirable at the beginning that the model is not too confident with respect to the segmentation.\ Too much confidence leads to the concentration of the sequence-level probability mass into fewer alignment paths.\  Early in training, these paths tend to be of low quality due to the random initialization of parameters.\ The uncertainty introduced by the blank label plays a crucial role in ensuring convergence.\ The scaling of label posterior and transition probabilities within the HMM topology plays a comparable role.\ The scales regulate the smoothness of the distribution, influencing the model's confidence in deciding whether to remain in the current label or transition to a different one~\cite{Zeyer+Beck+:Interspeech2017}.\ Previous work has shown also the relation between these scales and the input frame shift~\cite{raissi2024investigating}.\ Here, we argue that the contribution of the additional right and left factors in the gradient of \cref{eq:gradientsimplify} have also a role on the convergence, due to the inclusion of the output labels information on a wider range.\ %Experimental results in \cref{subsec:results} show the effect of the additional factors on the ASR accuracy.\ It is possible to see that the inclusion has a larger impact when less amount of data and training epochs are involved.

\subsection{Simplifying Assumptions}
\label{subsec:simplify}
\subsubsection{Context-independent factors}
The frame-wise label posteriors in the modern ASR approaches are modeled using neural networks.\ In locally normalized models, the final softmax layer defines a probability distribution over the set of acoustic model labels.\ Historically, due to data sparsity issues, the acoustic label units associated with triphones were defined through parameter tying, clustering similar triphones according to a predefined criterion~\cite{young1994tree}.\ Even under the assumption that all triphones are well represented in the data, training a model with a softmax layer of size equal to the phoneme set raised to the power of three, as the posterior used in \cref{eq:gradient}, is impractical.\ Recently, a factored hybrid HMM with no state-tying has been implemented and extensively evaluated~\cite{raissi2020fh,raissi2022hmm,raissi2023competitive,raissi2024investigating}.\ Factorization of the joint probability $P(\leftctx , a_{s_t},\rightctx| h_t)$ enables the original softmax layer parameters to be split into three distinct softmax layers, each normalized over the set of phonemes~\cite{raissi2020fh}.\ The triphone factored hybrid model is generally trained with frame-wise cross-entropy criterion with a fixed given alignment, known as Viterbi training.\ In order to calculate the original joint distribution, each of the network branches with context-dependent output needs to be forwarded with the given context embeddings and the output of all three softmax layers should be combined.\  The score computation for this model is generally more expensive in full-sum than in Viterbi.\ At a certain time frame the center phoneme of an alignment state could appear in different contexts on different alignment paths.\ Therefore, we propose a simplifying assumption for training that after the factorization of the joint probability in \cref{eq:gradient} drops the dependency to the phoneme contexts for each factor.\ The resulting gradient is the weighted sum of each $\partial \log$ probabilities of left, center, and left phonemes, where the weight for each $\partial \log$ is obtained by marginalizing the original joint $\gamma$ over the remaining phonemes, as shown in \cref{eq:gradientsimplify}.\ In our experiments, we observed that despite this simplifying assumption, given the input features, the three separate distributions for left, center, and right phonemes, learn the triphone structure implicitly.\ 

\subsubsection{Omitting the Prior During Training}
The scaled prior of \cref{eq:gradientsimplify} can be estimated on-the-fly during the training with exponential decaying average~\cite{Zeyer+Beck+:Interspeech2017} or as a fixed prior estimated on the transcriptions~\cite{Raissi+Zhou+:2023}.\ We observed convergence problems during training when using a prior and therefore considered the special case of prior scale $\beta=0$.\ Note that for the monophone HMM model that uses only the center phoneme state, setting the prior scale to zero effectively results in a posterior HMM~\cite{Raissi+Zhou+:2023}.

\section{Experimental Results and Setting}
\label{sec:exp}

\subsection{Setting}
\label{subsec:setting}
The experiments are carried out on 300h Switchboard-1 (SWB) Release 2 (LDC97S62) \cite{godfrey1992SWB} and 960h LibriSpeech (LBS) \cite{povey2015librispeech}.
We evaluate our models on SWB and CallHome subsets of Hub5`00 (LDC2002S09), the three subsets of Hub5`01 (LDC2002S13), as well as LBS dev and test sets.
The loss augmented with the right and left context is an extension of the CUDA based forward-backward within RETURNN~\cite{doetsch2017returnn}.\ Decoding of HMM based models use RASR for the core algorithms, and its recent extension for CTC decoding~\cite{rybach2011rasr,zhou2023rasr2}. Our experimental workflow is managed by Sisyphus~\cite{peter2018sisyphus}.
The speech signal is extracted using a 25ms window with a 10ms shift, yielding Gammatone filterbank features with dimensions of 40 for SWB and 50 for LBS~\cite{schluter2007Gammatone}.\ SpecAugment is applied across all models~\cite{park2019specaugment}. All encoder architectures consist of a 12-layer Conformer encoder with 75 million parameters~\cite{Zhou+Michel+:2022,gulati2020conformer}.\ The experiments of \cref{tab:onestage} are trained for 50 and 25 epochs for SWB and LBS, respectively.\ The total number of epochs is doubled for the experiments in \cref{tab:twostage}, also in case of experiments that used the seed models of \cref{tab:onestage} for initialization.\ We use one cycle learning rate schedule~(OCLR) up to peak LR of 6e-4 over 90\% of the training epochs, followed by a linear decrease to 1e-6~\cite{smith2019super,Zhou+Michel+:2022}.\ The Exp. 9 and 10 from \cref{tab:twostage} are initialized with the seed model and further trained with a constant learning rate of 5e-5 on 90\% of the training epochs followed by a linear decrease to 1e-6.\ An Adam optimizer with Nesterov momentum, together with optimizer epsilon of 1e-8 are used~\cite{dozat2016incorporating}.\ Decoding is performed using time-synchronous beam search based on dynamic programming principles with a lexical prefix tree.\ We use the official 4-gram LM for both tasks.\
%and an additional  transformer neural LM for LBS~\cite{beck2020lvcsr}.\
 For the real time factor measurement experiment, we used an AMD CPU (released 2021), with 2 logical cores.\ Example setups with more details are available.\footnote{\tiny{\url{https://github.com/rwth-i6/returnn-experiments/tree/master/2025-factored-fullsum-rightcontext}}}.
	\begin{table}[t]
	\setlength{\tabcolsep}{0.1em}\renewcommand{\arraystretch}{1.2}  % horizontal space inside cells.
	\centering \footnotesize %\captionsetup{margin=0cm,width=14cm}
	\caption{ASR performance models trained for 50 epochs on SWB300h and 25 epochs on LBS960h, using a 4-gram LM.\ We distinguish between the loss outputs used during training and the output used for decoding.\ The notation stands for a conditional probability of left, center, and right labels $\ell$, $c$, and $r$ given the encoder output $h$, respectively.}
	\label{tab:onestage}
	\begin{tabular}{|c||c|c|c||c|c||c|c|}
		\hline
		\multirow{2}{*}{\#} & \multirow{2}{*}{Model}  & \multicolumn{2}{c||}{Outputs} & \multicolumn{4}{c|}{WER {[}\%{]}}  \\ \cline{3-8} 
		& & Loss & Decode & \multicolumn{1}{c|}{Hub5`00}  &   \multicolumn{1}{c||}{Hub5`01}  & dev-other &test-other        \\ \hline
		1 & CTC & \multirow{2}{*}{$(c|h)$} & \multirow{3}{*}{$(c|h)$} & 12.8& 11.8& 6.9 &7.4 \\ \cline{1-2} \cline{5-8} 
		2& \multirow{3}{*}{HMM} &         &  &12.8 & 12.2 &6.9 & 7.3 \\  \cline{1-1} \cline{3-3}\cline{5-8} 
		3&									  & $(\ell|h)(c|h)(r|h)$ && \textbf{12.0} & \textbf{11.6}& \textbf{6.6}& 7.1 \\ \cline{1-1} \cline{3-8} 
		4&									  & $(\ell, c|h)$ & $(\ell, c| h)$ &12.4 & 11.9& \textbf{6.6}& \textbf{7.0} \\\hline		
	\end{tabular}
\end{table}
\begin{table}[t]
	\setlength{\tabcolsep}{0.01em}\renewcommand{\arraystretch}{1.2}  % horizontal space inside cells.
	\centering \footnotesize  %\captionsetup{margin=0cm,width=14cm}
	\caption{Similar experiments as in \cref{tab:onestage} with double amount of epochs.\ Two diphone Exps. 9 and 10 are initialized with the models from Exps. 2 and 3. The total number of epochs is the same as all other models in the table.\ }
	\label{tab:twostage}
	\begin{tabular}{|c|c|c|c|c||c|c||c|c|}
		\hline
		\multirow{2}{*}{\#} & \multirow{2}{*}{Model}  & \multicolumn{2}{c|}{Outputs} &  \multirow{2}{*}{Init}  & \multicolumn{4}{c|}{WER {[}\%{]}}  \\  \cline{3-4} \cline{6-9} 
		& & Loss & Decode & &  \multicolumn{1}{c|}{Hub5`00}  &   \multicolumn{1}{c||}{Hub5`01}  & dev-other &test-other        \\ \hline
		5 & CTC & \multirow{2}{*}{$(c|h)$} & \multirow{3}{*}{$(c|h)$} &   \multirow{4}{*}{-}&12.3 & 11.3& 6.6& 6.9 \\ \cline{1-2} \cline{6-9} 
		6& \multirow{6}{*}{HMM}    &      &  && 12.2 & 11.7 & 6.3 & 6.6 \\  \cline{1-1} \cline{3-3}\cline{6-9} 
		7&	&			 $(\ell|h)(c|h)(r|h)$ & & & \textbf{11.6}& 11.0&6.1 & 6.5  \\ \cline{1-1} \cline{3-4 }\cline{6-9} 
		8&	& \multirow{3}{*}{$(\ell, c|h)$} & \multirow{3}{*}{$(\ell, c| h)$} &&11.8 & 11.0& \textbf{6.0}& \textbf{6.4} \\ \cline{1-1} \cline{5-9} 
		9&&&&\#2&11.8 & 10.9& 6.3& 6.7 \\   \cline{1-1} \cline{5-9} 
		10&&&&\#3&\textbf{11.6} & \textbf{10.8}& 6.2& 6.8 \\ \hline		
	\end{tabular}\vspace{-0.5cm}
\end{table}

\begin{table}[t]
	\setlength{\tabcolsep}{0.5em}\renewcommand{\arraystretch}{1.0}  % horizontal space inside cells.
	\centering \footnotesize 
	\caption{Comparisons of training time and real time factor~(RTF) for two experiments from \cref{tab:twostage} on dev-other.}
	\label{tab:rtf}
	\begin{tabular}{|c|c|c||c||c|c|} 
		\hline			
		\multirow{2}{*}{\#} & \multicolumn{2}{c||}{ \multirow{1}{*}{Outputs}} & \multirow{2}{*}{Train[h]} &  \multirow{2}{*}{RTF} & WER [\%] \\ \cline{2-3} \cline{6-6}
		&  Loss& Decode & && dev-other \\ \hline
		7 &  $(\ell|h)(c|h)(r|h)$  & $(c|h)$ & \textbf{300} & \textbf{0.099} & 6.1\\ \hline
		8 &  $(\ell,c|h)$  & $(\ell,c|h)$ & 330 & 0.118& \textbf{6.0}\\ \hline
	\end{tabular}  
\end{table}

\begin{table}[t]	
	\setlength{\tabcolsep}{0.01em}\renewcommand{\arraystretch}{1.0}  % horizontal space inside cells.
	\centering \footnotesize 
	\caption{Diphone factored and monophone hybrid models on LBS 960h decoded with 4-gram LM.\ The external alignment~(ext. align) is a from-scratch posterior HMM with BLSTM encoder trained for 25 epochs.\ The Viterbi training for the remaining 20 epochs relies on factored outputs without the simplifying assumption.\ The other two models use only full-sum and do not use any external alignment. }
	\label{tab:scratch}
	\begin{tabular}{|c|c|c|c||c|c|} 
		\hline
		 Ext.& \multicolumn{3}{|c||}{Train} &  \multicolumn{2}{c|}{WER [\%]} \\ \cline{2-6}
		 Align &Loss & Decode & Epochs& dev-other &test-other \\ \hline
		  yes & $(\ell|h)(c|\ell, h)$&  \multirow{2}{*}{ $(\ell,c|h)$} &45 & 6.1& 6.8 \\ \cline{1-2}\cline{4-6}
		 \multirow{2}{*}{no} &$(\ell,c|h)$& & \multirow{2}{*}{50}  & \textbf{6.0}& \textbf{6.4} \\ \cline{2-3} \cline{5-6}
		 &$(\ell|h)(c|h)(r|h)$ & $(c|h)$ &  & 6.1 &6.5\\ \hline
	\end{tabular}  \vspace{-0.5cm}
\end{table}

\subsection{Results}
\label{subsec:results}

We start with the evaluation of a first set of models trained from scratch in \cref{tab:onestage}.\ All models are trained for the same number of epochs, i.e., 50 epochs for SWB 300h and 25 epochs for LBS 960h.\ It is possible to see that among the zero order label context models, the inclusion of the right context in Exp. 3 yields the best results for SWB task.\ The performance gap is smaller when using LBS.\ However, model of Exp. 3 still outperforms all other zero-order label context models.\ We then evaluate the effect of longer training and the model initialization in \cref{tab:twostage}.\ Here, we use the two models from Exps. 2 and 3 to initialize the diphone factored hybrid that is trained for additional 50 and 25 epochs for SWB and LBS, respectively.\ All models in \cref{tab:twostage} are trained for the same number of epochs, also when they are trained without initialization.\ The diphone model can take advantage of the pretrained model of Exp. 3 on SWB 300.\ This does not seem to be the case for LBS task where the best performance is obtained by the from-scratch diphone model.\ However, the model of Exp. 7 with the right context is also performing in the same range as the diphone of Exp. 8.\ In \cref{tab:rtf}, we take a closer look at the training time and real time factor of the monophone HMM with left and right auxiliary losses (Exp. 7) and diphone factored hybrid (Exp. 8).\ The model in Exp. 7 shows a $1.6\%$ relative increase of WER, which is negligible considering the $10\%$ relative speedup in training time and $16\%$ relative improvement in RTF.\ Furthermore, comparison between the Exps. 1-4 of \cref{tab:onestage} and the Exps. 5-8 of \cref{tab:twostage} highlights the advantage of the inclusion of the right label context even when training for fewer epochs.\ Moreover, in all experiments we observed that HMM based models outperformed the CTC under similar conditions.\

In our previous works, we proposed a Viterbi trained diphone factored hybrid using the alignment from a monophone posterior HMM, with comparable or superior performance to a strictly monotonic RNN-T~\cite{raissi2023competitive,raissi2024investigating}.\ As a further step, in \cref{tab:scratch} we show that our proposed end-to-end training can lead to similar ASR accuracy as the standard hybrid HMM pipeline, bringing our HMM based systems closer to the end-to-end approaches, in terms of simplicity.

\section{Conclusions}
We showed that the inclusion of right label context in training of time-synchronous ASR models lead to performance gain particularly when data resources are limited.\ We also proposed an end-to-end training solution for hybrid HMM systems eliminating the need for the standard complex multi-stage pipeline and the external alignments while still achieving comparable performance.\

\normalsize 
\section{Acknowledgments} \footnotesize
This work was partially supported by NeuroSys, which as part of the initiative “Clusters4Future” is funded by the Federal Ministry of Education and Research BMBF (03ZU1106DA). The authors thank Albert Zeyer for the invaluable discussions and Eugen Beck for their unwavering support of the factored hybrid concept.

\bibliographystyle{IEEEtran}
\bibliography{references}

\end{document}